\documentclass[aps,prc,twocolumn,showkeys,showpacs]{revtex4}
\usepackage{amssymb}
\usepackage{graphicx}
\usepackage{multirow}
\usepackage{color}

\def\bra{\langle}
\def\ket{\rangle}

\begin{document}

\title{To fission or not to fission}
\author{K. Pomorski}
\email{Krzysztf.Pomorski@umcs.pl}
\affiliation{Maria Curie Sk\l odowska University, Lublin, Poland}
\author{F.A.~Ivanyuk}
\email{ivanyuk@kinr.kiev.ua}
\affiliation{Institute for Nuclear Research, Kiev, Ukraine}
\author{B. Nerlo-Pomorska}
\email{pomorska@kft.umcs.lublin.pl}
\affiliation{Maria Curie Sk\l odowska University, Lublin, Poland}
\date{\today}


\begin{abstract}
The fission-fragments mass-yield of $^{236}$U is obtained by an approximate 
solution of the eigenvalue problem of the collective Hamiltonian that describes 
the dynamics of the fission process whose degrees of freedom are: the fission 
(elongation), the neck and the mass-asymmetry mode. The macroscopic-microscopic 
method is used to evaluate the potential energy surface. The macroscopic energy 
part is calculated using the liquid drop model and the microscopic corrections 
are obtained using the Woods-Saxon single-particle levels. The four dimensional 
modified Cassini ovals shape parametrization is used to describe the shape of 
the fissioning nucleus. The mass tensor is taken within the cranking-type 
approximation. The final fragment mass distribution is obtained by weighting the 
adiabatic density distribution in the collective space with the neck-dependent 
fission probability. The neck degree of freedom is found to play a significant 
role in determining that final fragment mass distribution.
\end{abstract}
\maketitle

\section{Introduction}

A very stringent test of any theoretical model which describes the nuclear 
fission process should be a proper reproduction of the fission fragments mass 
distribution. The goal of the present paper is to obtain such a distribution by 
an approximate solution of the eigenvalue problem of the 3-dimensional 
collective Hamiltonian with degrees of freedom corresponding to elongation, neck 
formation and mass asymmetry of the nuclear shape. The present model is similar 
to the 2-dimensional one of Refs. \cite{Berlin,APPBS} but the non-adiabatic and 
dissipative effects are not taken here into account since their effect is rather 
small for low-temperature fission. The potential energy surface (PES) is 
obtained in the present work using the macroscopic-microscopic method with the 
liquid drop model for the macroscopic part of the energy while the microscopic 
shell and pairing corrections are calculated using the Woods-Saxon (WS) 
single-particle levels \cite{pash71}. The shape of the fissioning nucleus is 
described by the four dimensional modified Cassini ovals (MCO) 
\cite{pash71,pasrus}. It was shown in Ref. \cite{caivpa} that the MCO describe 
very well the so-called optimal nuclear shapes obtained through a variational 
description \cite{IvaPom}, even those close to the scission configuration. The 
mass tensor is taken within the cranking-type approximation (confer e.g. Sec. 
5.1.1 of Ref.~\cite{KP12}). The Born-Oppenheimer approximation (BOA) is used to 
describe the coupling of the fission mode with the neck and mass asymmetry 
degrees of freedom. It will be shown that, in order to obtain a fission-fragment 
mass distribution in agreement with the experimental data, that fission 
probability should depend on the neck size.

The paper is organized in the following way. First we shortly present the 
details of our theoretical model, then we show the collective potential energy 
surface  evaluated in the macroscopic-microscopic model for $^{236}$U and the 
components of the mass tensor. The calculated fission fragments mass 
distribution is compared with the experimental data in the next section. 
Conclusions and possible extensions and applications of our model are presented 
in Summary.

\section{Collective Hamiltonian}
\subsection{Shape parameterization}

\begin{figure}[ht] 
\centerline{\includegraphics[width=0.49\textwidth]{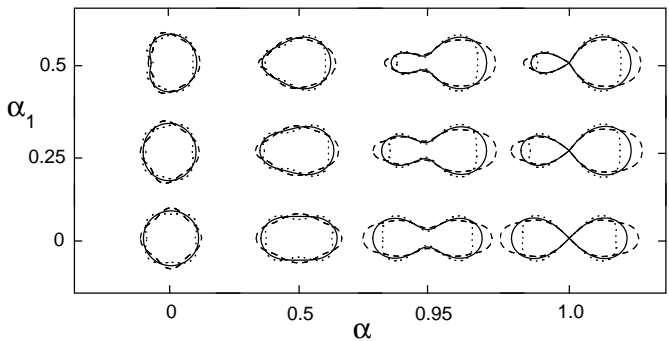}}
\caption{Examples of nuclear shapes in $\alpha, \alpha_1, \alpha_4$ parametrization. 
The solid lines correspond to $\alpha_4$=0 while the dashed and dotted curves 
to $\alpha_4$=0.2 and $\alpha_4$=-0.2 respectively.} 
\label{cassini}
\end{figure}

We define the shape of fissioning nucleus by the
parameterization developed in \cite{pash71}. In this
parameterization some cylindrical co-ordinates \{$\overline{\rho}$, $\overline{z}$\} are related to the lemniscate co-ordinates system $\{R, x\}$ by the equations
\begin{eqnarray}\label{eq:LSC}
\overline{\rho}&=&\frac{1}{\sqrt{2}}\sqrt{p(x)-R^2(2x^2-1)-s}\,,\, \nonumber\\
\overline{z}&=&\frac{{\rm sign}(x)}{\sqrt{2}}\sqrt{p(x)+R^2(2x^2-1)+s},
\nonumber \\
p^2(x)&\equiv& R^4+2sR^2(2x^2-1)+s^2, \nonumber\\
&&0\leq R\leq \infty, -1\leq x \leq 1.
\end{eqnarray}
The co-ordinate surfaces of the lemniscate system $R(x)=R_0$ are the Cassini ovals (see the bottom of Fig. \ref{cassini})
with $s\equiv \varepsilon R_0^2$, where $s$ is the squared distance between the focus of
Cassinian ovals and the origin of coordinates. 
%%%%%%%%%%%%%%%%%%%%%%%%%%%%%%%%%%%%%%%%%%%%%%%%%%%%%%%%%%%%%%%%%%%%%%%%%%%%%%
%%%%%%%%%%%%%%%%%%%%%%%%%%%%%%%%%%%%%%%%%%%%%%%%%%%%%%%%%%%%%%%%%%%%%%%%%%%%%%

The deviation of the nuclear surface from Cassini ovals is defined by expansion of $R(x)$
in series in Legendre polynomials $P_n(x)$,
\begin{equation}\label{Rx}
R(x)=R_0[1+\sum _n \alpha _n P_n(x)]\,,
\label{eq:Pol}
\end{equation}
where $R_0$ is the radius of the spherical nucleus.   The cylindrical co-ordinates $\{{\rho}, {z}\}$ are related to  $\{\overline{\rho}, \overline{z}\}$ by
\begin{equation} 
\rho\equiv\overline{\rho}/c,\,\quad z\equiv(\overline{z}-
\overline{z}_{\rm cm})/c\,,
\end{equation}
where  $\overline{z}_{\rm cm}$ is the $z$-coordinate of the center of mass of 
Cassini ovaloid (\ref{Rx}) and the constant $c$ is introduced in order to insure the volume conservation.

Instead of $\varepsilon$, it turns out convenient to introduce another 
parameter, $\alpha$, which is defined so, that at $\alpha=1$ the neck radius 
turns into zero for any value of all other deformation parameters $\alpha _n$,
\begin{eqnarray}
\varepsilon=\frac{\alpha-1}{4}[(1+\sum _n \alpha _n)^2+(1+\sum _n (-1)^n \alpha _n)^2] \nonumber\\
+\frac{ \alpha+1}{2}[1+\sum _n (-1)^n \alpha _{2n}(2n-1)!!/(2^nn!)]^2.\label{eq:def}
\end{eqnarray} 
The parameters $\alpha$ and $\alpha _n$ are considered as the deformation parameters.
Examples of the shapes in Cassini parameterization are shown in Fig. \ref{cassini}.
%%%%%%%%%%%%%%%%%%%%%%%%%%%%%%%%%%%%%%%%%%%%%%%%%%%%%%%%%%%%%%%%%%%%%%%%%%%%%%%%%%%%%
\subsection{Born-Oppenheimer approximation}
%%%%%%%%%%%%%%%%%%%%%%%%%%%%%%%%%%%%%%%%%%%%%%%%%%%%%%%%%%%%%%%%%%%%%%%%%%%%%%%%%%%%%%
Below we use the following collective coordinates to describe the fission 
dynamics:
\begin{equation}
    q^1=R_{12}~,~~~
    q^2=\mathfrak{a}=\frac{{\cal V}_{1}-{\cal V}_{2}}{{\cal V}_{1}+{\cal V}_{2}}
    ~,~~{\rm and}~~ q^3=\alpha_4\,.
\label{cc}
\end{equation}
Here $R_{12}$ is the distance between the mass center of the nascent fragments 
in units of the radius $R_0$ of the corresponding spherical nucleus, while $ 
\mathfrak{a}$ is the mass asymmetry coordinate. ${\cal V}_1$ and ${\cal V}_2$ 
are the volumes of the fragments and $\alpha_4$ describes the neck degree of 
freedom when $R_{12}$ is kept constant (see Ref.\cite{pasrus}).\\
With these coordinates the classical energy of the system becomes
\begin{equation}
H_{\rm cl}=\frac{1}{2}\sum _{i, j} M_{ij} \dot q^i\dot q^j+
V(\{q^i\}) ~,
\label{H=}
\end{equation}
where $M_{ij}$ and $V(\{q_i\})$ denote the inertia tensor and the potential
energy, respectively.\\
The quantized form of this Hamiltonian is the following:
\begin{equation}
\widehat H=-\frac{\hbar^2}{2} \sum_{i,j} |M|^{-1/2} \frac{\partial}{\partial
q^i}
      |M|^{-1/2} M^{ij}\frac{\partial}{\partial q^j} + V(\{q^i\})~,
\end{equation}
where  $|M|=\det(M_{ij})$ and $M_{ij}M^{jk}=\delta^{k}_{i}$.

The eigenproblem of this Hamiltonian will be solved here in the
Born-Oppenheimer approximation in which one assumes that the motion towards
fission is much slower than the one in the two other collective coordinates. In
such an approximation the Hamiltonian could be written as follows:
\begin{equation}
\hat H(R_{12},\mathfrak{a},\alpha_4)\approx \hat T_{\rm fis}(R_{12}) +
\hat H_{\rm perp}(\mathfrak{a},\alpha_4;R_{12}) \,.
\label{hadia}
\end{equation}
Here $\hat T_{\rm fis}$ is the fission mode kinetic energy operator
\begin{equation}
\hat T_{\rm fis}(R_{12})=-\,\frac{\hbar^{2}}{2}\,
\frac{\partial}{\partial R_{12}}\,\overline M^{-1}(R_{12})\,\frac{
\partial}{\partial R_{12}}\,,
\label{Tfis}
\end{equation}
where $\overline{M}(R_{12})$ is the average inertia related to the fission mode
and $\hat H_{\rm perp}$ is the collective Hamiltonian related to the neck and 
mass asymmetry coordinates. The eigenfunction of Hamiltonian (\ref{hadia}) can 
be written as a product:
\begin{equation}
\Psi_{nE}(R_{12},\mathfrak{a},\alpha_4)=
                   u_{nE}(R_{12})\varphi_{n}(\mathfrak{a},\alpha_4;R_{12})~,
\end{equation}
where $\varphi_n$ are the eigenfunctions of $\hat H_{\rm
perp}$:
\begin{equation}
\hat H_{\rm perp}\,
\varphi_{n}(\mathfrak{a},\alpha_4; R_{12})=
           e_n(R_{12}) \varphi_n(\mathfrak{a},\alpha_4;R_{12})~,
\label{hperp}
\end{equation}
and they are evaluated for each mesh-point value in the $R_{12}$ direction.
Using the above relations one can rewrite the eigenequation of the Hamiltonian
(\ref{hadia}) in the following form:
\begin{equation}
\left(\hat T_{\rm fis}+e_{n}(R_{12})\right)\,u_{nE}(R_{12})=
        E\,u_{nE}(R_{12})~.
\label{Had}
\end{equation}
The approximate solution of the above eigenvalue problem can be obtained using 
the WKB formalism. The energies $e_n(R_{12})$ in Eq. (\ref{Had}) define the 
fission potential for different channels, what is important when one describes 
the nonadiabatic fission process in the coupled channels approach \cite{APPBS}. 
In the following we shall take only the lowest energy channel, what corresponds 
to the adiabatic approximation. Within this approximation the wave function of 
the fissioning nucleus is written in the form of a product of the wave function 
$u_{0E}(R_{12})$ describing the motion towards fission and the function 
$\varphi_0(\mathfrak{a},\alpha_4;R_{12})$ which corresponds to the lowest 
eigenenergy $e_0$ of the Hamiltonian (\ref{hperp}). The probability of finding 
of a nucleus, for a given value of $R_{12}$, in the given 
($\mathfrak{a},\alpha_4$) point is equal to 
$|\varphi_0(\mathfrak{a},\alpha_4;R_{12})|^2$.

In Refs.~\cite{Berlin,APPBS} is shown how to go beyond the BOA and include
nonadiabatic and dissipative effects, but in the following we are going to omit 
such effects, which are expected to be small at low temperatures and limit our 
discussion to the effect of the neck degree of freedom on the fission-fragment 
mass distribution.

%----------------------------------------------------------------------------

\section{Numerical results}
\label{sec:present}

The potential energy surface is evaluated for $^{236}$U at zero temperature
within the macroscopic-mic\-ros\-co\-pic model in which the macroscopic part of
the energy is obtained using the liquid drop formula and the microscopic shell
and pairing corrections are calculated using the Woods-Saxon single-particle
potential. All parameters of the calculation are described in
Ref.~\cite{pasrus}. The potential energy was calculated in 4-dimensional space
of deformation parameters $\alpha, \alpha_1, \alpha_4, \alpha_6$ and
minimized then with respect to $\alpha_6$.

The mass parameter $M_{ij}(q)$ for the fission process is commonly  calculated by the Inglis formula
\begin{equation}\label{inglis}
M_{ij}(q)= 2\hbar^2 \sum_m \frac{\bra 0\vert\partial/\partial q^i\vert 
m\ket\bra m\vert\partial/\partial q^j\vert 0\ket}{E_m-E_0}\,,
\end{equation}
where $\vert 0\ket$ and $\vert m\ket$ denote the ground and an excited state of the system. 
%%%%%%%%%%%%%%%%%%%%%%%%%%%%%%%%%%%%%%%%%%%%%%%%%%%%%%%%%%%%%%%%%%%%%%%%%%%%%%%
\begin{figure*}[!htb]
\includegraphics[width=0.98\textwidth]{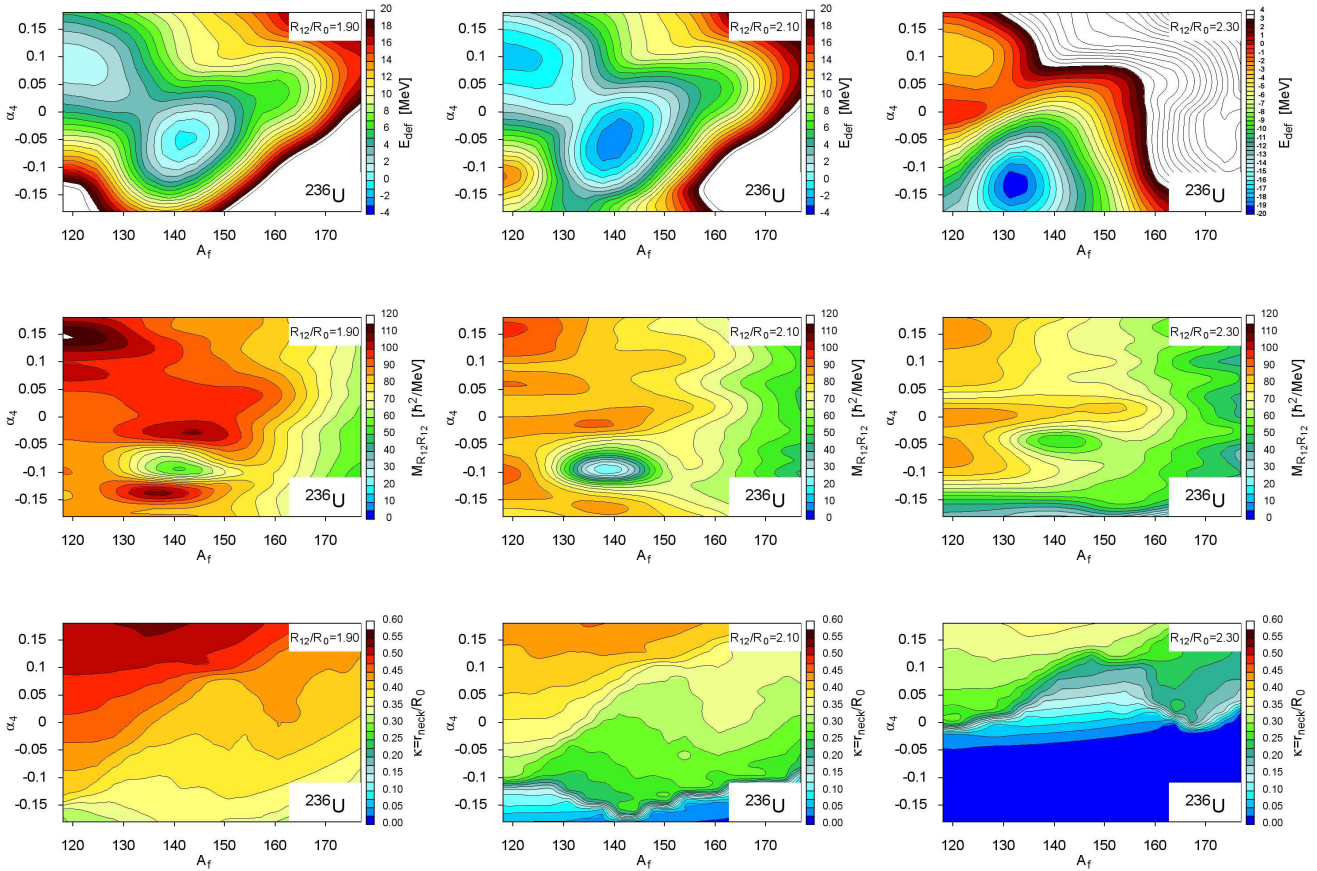}
\caption{(Color online) Macroscopic-microscopic PES (upper row) and the
relative distance component $M_{R_{12}R_{12}}$ of the inertia tensor (middle 
row) and the neck radius  $\kappa=r_{\rm neck}/R_0$ (lower row) on the 
$(A_{\rm f},\alpha_4)$ plane for different values of the elongation $R_{12}$.}
\label{evolution}
\end{figure*}
%%%%%%%%%%%%%%%%%%%%%%%%%%%%%%%%%%%%%%%%%%%%%%%%%%%%%%%%%%%%%%%%%%%%%%%%%%%%%%%

In the case that the ground and the excited states of the system are described 
by the BCS approximation, the $M_{ij}(q)$ is given by 
\cite{brdapa},
\begin{equation}\label{bij}
M_{ij}=2\hbar^2\sum_{\mu\nu}\frac{\bra\mu\vert\partial H/ \partial 
q^i\vert\nu\ket
\bra\nu\vert\partial H/ \partial q^j\vert\mu\ket}{(E_{\mu}+E_{\nu})^3} 
\eta_{\mu\nu}^2 +P_{ij}\,,
\end{equation}
where the term $P_{ij}$
stands for the contribution due to the change of occupation numbers, when the 
deformation varies. The $E_{\mu}$, $u_{\mu}$ and $v_{\mu}$ in (\ref{bij}) are 
the quasi-particle energies and coefficients of the Bogolyubov-Valatin 
transformation correspondingly, and $\eta_{\mu\nu}\equiv 
u_{\mu}v_{\nu}+u_{\nu}v_{\mu}$.

Unfortunately, the expression (\ref{bij}) has the very unpleasant feature that 
it does not turn into the mass parameter of a system of independent particles 
when the pairing vanishes, $\Delta\to 0$. More precisely, the non diagonal sum 
over single-particle states in (\ref{bij}) does turn into the mass parameters of 
the system of independent particles when $\Delta\rightarrow 0$, but, even worse, 
the diagonal sum goes to infinity in that limit (it is proportional 
$1/\Delta^2$, as demonstrated in \cite{brdapa}). Thus, at some points in the 
deformation space, where the density of single-particle states is very low, the 
mass parameter (\ref{bij}) becomes unreasonably large. The same happens in 
excited systems, when the temperature is close to it's critical value $T_{crit}$ 
at which the pairing gap disappears.

One should also keep in mind that the diagonal contribution to the sum in 
(\ref{bij}) comes from the matrix elements between the ground state and the pair 
excited states that correspond to the particle number, different from that of 
the ground state. In a particle number conserving theory such contribution could 
not appear.

In order to avoid the problems related to the diagonal contribution to 
(\ref{bij}) we have omitted in (\ref{bij}) the diagonal 
matrix elements $\bra\mu\vert\partial H/ \partial q_i\vert\mu\ket$ and taken 
into account only the non-diagonal part of (\ref{bij}),
\begin{equation}\label{bij_nd}
M_{ij}=2\hbar^2\sum_{\mu}\sum_{\nu\neq\mu}\frac{\bra\mu\vert\partial H/ \partial 
q^i\vert\nu\ket
\bra\nu\vert\partial H/ \partial q^j\vert\mu\ket}{(E_{\mu}+E_{\nu})^3} 
\eta_{\mu\nu}^2\,.
\end{equation}
The inertia tensor (\ref{bij_nd}) is evaluated in the 3-dimensional space of 
deformation parameters $\alpha, \alpha_1, \alpha_4$. For each value of 
$\alpha_4$ and the potential energy and the components of mass tensor were 
transformed from $\alpha, \alpha_1$ to $R_{12}, \mathfrak{a}$ coordinates 
defined in Eq.~(\ref{cc}). The potential energy surface (PES) related to the 
spherical liquid drop energy (upper row) and the $M_{\rm R_{12}R_{12}}$ 
component of the inertia tensor (middle row) as well as the neck radius 
$\kappa=r_{\rm neck}/R_0$ (lower row) are plotted in Fig.~\ref{evolution} on the 
$(A_{\rm f},\alpha_4)$ plane for three different values of the relative 
distances of the center of the fragments $R_{12}$. Here $A_{\rm f}$ is the 
atomic mass of fission fragment.

In the following we would like to describe the way in which one has
to obtain the fission fragment mass yield after solving the quantum mechanical
problem of the collective Hamiltonian which describes the fission process in
the three dimensional space (3D) composed of the following deformation
parameters:\\[-4ex]
\begin{description}
\item[$R_{12}$] -- distance between the mass center of the fission
fragments,\\[-4ex]
\item[$\mathfrak{a}$] = ${[A_{\rm f}(1)-A_{\rm f}(2)]}/{[A_{\rm
f}(1)+A_{\rm f}(2)]}$ -- the mass asymmetry
     parameter, where $A_{\rm f}(1)$ and $A_{\rm f}(2)$ are the mass numbers of
the fission
     fragments,\\[-4ex]
\item[$\alpha_4$] -- hexadecapole correction to the Cassini ovals
\cite{pash71}.\\[-4ex]
\end{description}

First one has to prepare a set of the 2D mass distributions
\begin{equation}
|\Psi(\mathfrak{a},\alpha_4;R_{12})|^2 = 
|\varphi_0(\mathfrak{a},\alpha_4;R_{12})|^2
\label{dipo}
\end{equation}
on the plane $(\mathfrak{a},\alpha_4)$
by solving eigenproblem of a corresponding 2D collective Hamiltonian for fixed
elongations $R_{12}$ \cite{Nerlo}. One has to bear in mind that it is very
unlikely, that fission occurs at some fixed $R_{12}$ or when the system 
reaches the scission line/surface. The problem is much more
complicated and one has to involve into consideration the size of the neck.

\begin{figure}[!htb]
\begin{center}
\includegraphics[width=0.35\textwidth]{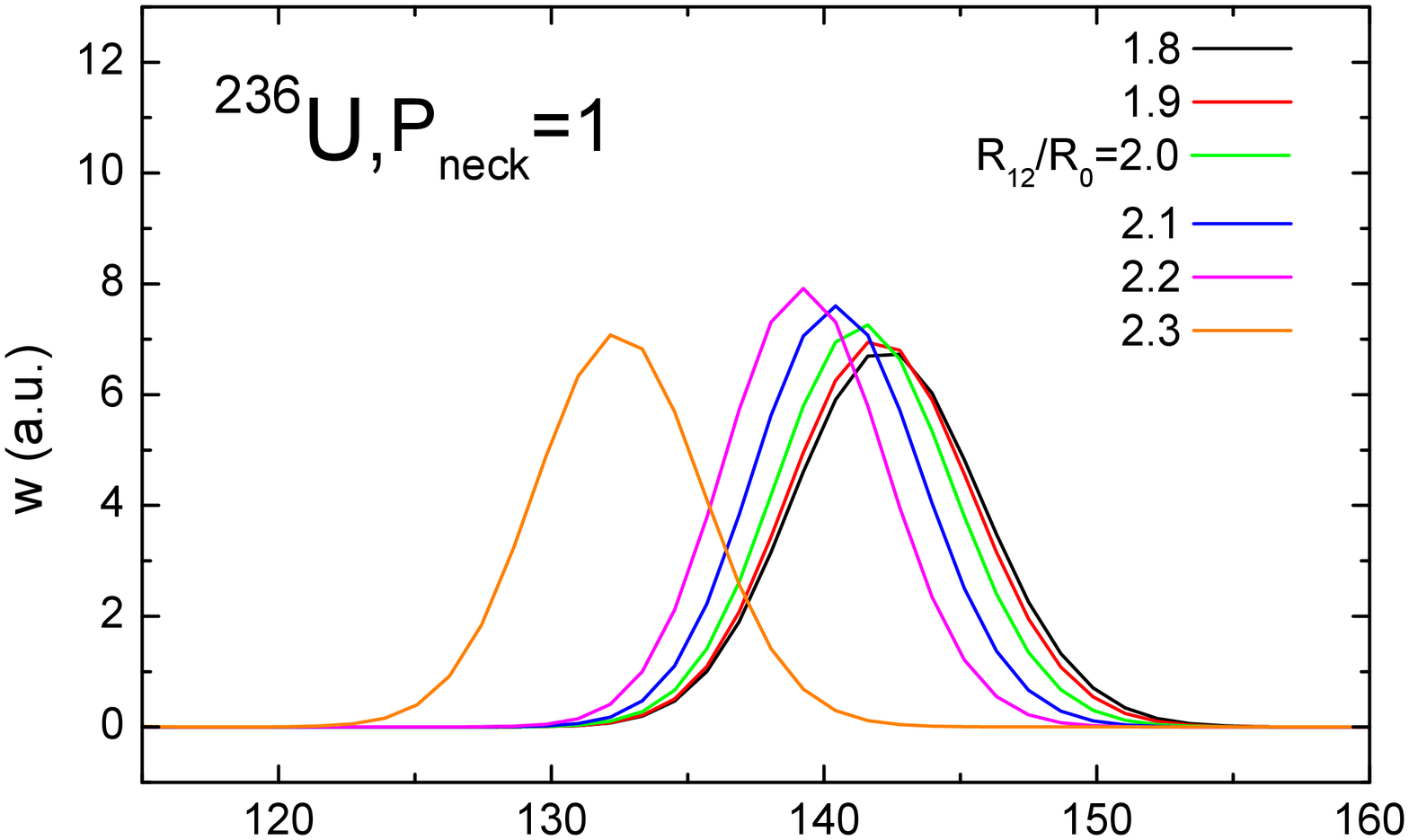}\\[1ex]
\includegraphics[width=0.35\textwidth]{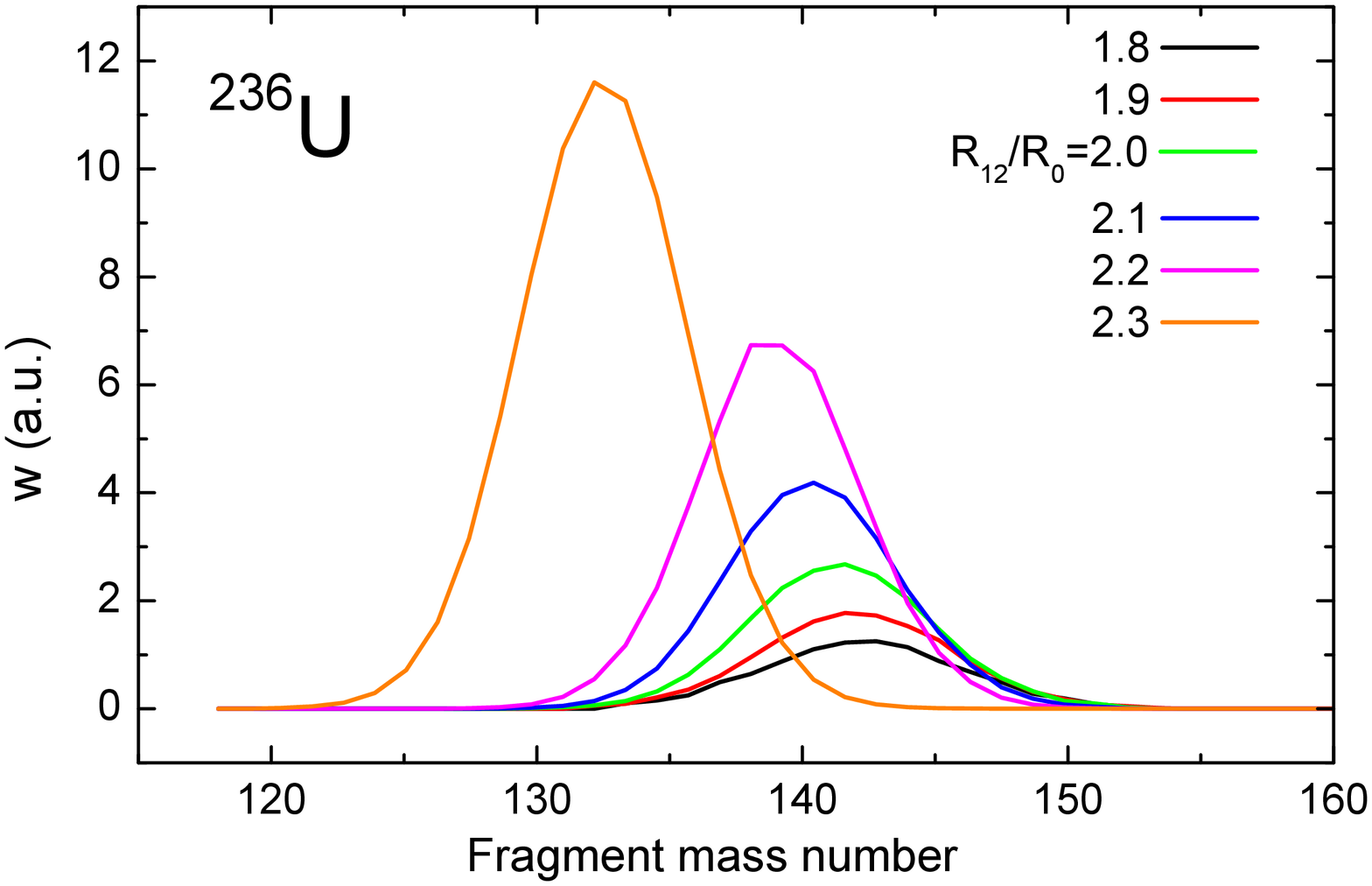}
\end{center}\vspace{-5mm}
\caption{(Color online) Probability distribution (\protect\ref{waR}) (top)
and the fission probability (\protect\ref{pda}) (bottom) calculated at few
fixed elongations of the fissioning nucleus.}
\label{2Dmass}
\end{figure}

Looking at the integrated over $\alpha_4$  probability distributions for
$^{236}$U presented in the top part of
Fig.~\ref{2Dmass}
\begin{equation}
   w(\mathfrak{a},R_{12})=\int |\Psi(\mathfrak{a},\alpha_4;R_{12})|^2 d\alpha_4~,
\label{waR}
\end{equation}
one can not see any qualitative change with respect to the results which we
have published in Ref. \cite{APPBS} for the calculation made in the
$(R_{12},\mathfrak{a})$ plane. Both distributions, i.e. the one corresponding
to $A_{\rm f}(1)=140$ for smaller $R_{12}$ and the one for $A_{\rm f}(1)=132$ at
$R_{12}$ close to the scission line are only slightly broader. So, the problem
to reproduce the experimental distribution of fission fragments, seen in the 2D 
space \cite{APPBS}, remains also in the 3D space.
%%%%%%%%%%%%%%%%%%%%%%%%%%%%%%%%%%%%%%%%%%%%%%%%%%%%%%%%%%%%%%%%%%%%%%%%%%%%%%%
%%%%%%%%%%%%%%%%%%%%%%%%%%%%%%%%%%%%%%%%%%%%%%%%%%%%%%%%%%%%%%%%%%%%%%%%%%%%%%%
The only solution is to assume that fission occurs with a certain
probability before (or after) reaching the critical elongation
$R_{12}^{crit}$. Depending on the neck radius a fissioning nucleus has
to make its choice ``{\it to fission or not to fission}''. When it
{\it decides for fission} it would leave the phase-space of collective
coordinates. Of course this is not a {\it Hamlet dilemma}, where there is
only the choice between {\it yes} or {\it no}. We are rather faced here
with a statistical problem and the answer {\it yes} is given with a certain
probability which one then will have to take into account in
the distribution probability (\ref{dipo}) in the phase space.

Following such an assumption a part of the events (read trajectories in the 
Langevin approach, or distributions in our quantum mechanical model) disappears 
from the phase-space and leads to a kind of weighting of the mass distribution 
corresponding to the different elongations $R_{12}$. To do this one has to 
evaluate the neck radius in the whole 3D space. The neck radius parameter 
$\kappa=r_{\rm neck}/R_0$ is plotted in the lower row  of Fig.~\ref{evolution} 
on the $(A_{\rm f},\alpha_4)$ plane for three different values of the 
elongation 
parameter $R_{12}$. The slight wiggles in Fig. \ref{evolution} are caused by the 
approximate minimization  with respect to the $\alpha_6$ deformation parameter. 
It is seen that on average the neck radius decreases with growing $R_{12}$. For 
a constant $R_{12}$ the neck radius varies strongly with $\alpha_4$ and 
$A_{\rm f}$. One commonly agrees that fission takes place when the neck 
radius becomes of the order of the size of a nucleon. 
This is the case for $\kappa\approx 0.2$, which is realized at 
$R_{12}/R_0=2.0$ for $\alpha_4=-0.18$; $R_{12}/R_0=2.25$ for $\alpha_4=0$, and 
$R_{12}/R_0>2.5$ for $\alpha_4=0.18$ and the asymmetry parameter 
$\mathfrak{a}\approx 0.2$.  From the optimal shape approach \cite{IvaPom} one 
knows that the scission shape corresponds to $r_{\rm neck}\approx 0.3R_0$ and 
$R_{12}^{\rm crit}\approx 2.3 R_0$. In the case of the Cassini parametrization 
used in the present paper the $r_{\rm neck}$ can be somewhat smaller.

One could try to parametrize the neck-rupture probability $P$ in the following
form:
\begin{equation}
      P(\mathfrak{a},\alpha_4,R_{12})=\frac{k_0}{k} P_{\rm 
neck}(\kappa)\,,
\label{pn}
\end{equation}
where $k$ is the momentum in the direction towards fission (or simply the 
velocity along the elongation coordinate $R_{12}$), while 
$\kappa=\kappa(\mathfrak{a},\alpha_4,R_{12})$ is the deformation dependent 
relative neck size. The scaling parameter $k_0$, plays no essential role, and 
will disappear from the final expression of the mass distribution when one 
will normalize it. The geometry dependent part of the neck breaking probability 
is taken in the form of a Fermi function:
\begin{equation}
      P_{\rm neck}(\kappa)=\left(1+e^\frac{\kappa-\kappa_0}{d}\right)^{-1}~.
\label{pneck}
\end{equation}
The parameters $\kappa_0$ and $d$ have to be fixed by comparing the theoretical
fission fragment mass distributions with the experimental ones. Our goal 
is to fix these parameters in a kind of universal way, independent of the 
specific fission reaction that one wants to investigate. The present 
investigation has to be treated only as a first attempt in this direction.

The momentum $k$ which appears in the denominator of Eq.~(\ref{pn}) has to 
ensure that the probability depends on time in which one crosses the subsequent 
intervals in $R_{12}$ coordinates: $\Delta t= \Delta R_{12}/v(R_{12})$,  where 
$v(R_{12})=\hbar k/\overline M(R_{12})$ is the velocity towards fission. The 
value of $k$ depends on the difference $E-V(R_{12})$ and on the part of the 
collective energy which is converted into heat $Q$:
\begin{equation}\label{kk}
\frac{\hbar^2 k^2}{2\overline M(R_{12})}=E_{kin}=E-Q -
        V(R_{12})~.
\end{equation}

In the quantum mechanical picture the heat $Q$ can be replaced by the imaginary 
part of the collective potential \cite{APPBS}. In the Langevin picture the 
method should be almost the same but one has also to work at least in the 3D 
space. In our present calculations we have put $Q=0$, i.e., we assumed a 
"complete acceleration" scenario: no dissipation takes place, which is 
reasonable since at low excitation energies the friction force is very week.

The $\overline M$ in (\ref{kk}) is the cranking inertia relative to the $R_{12}$ 
deformation parameter. In principle, we used the definition of cranking inertia, 
but for the reasons explained above the contributions from the diagonal matrix 
elements were removed.

The fission probability $w$ at a given $R_{12}$ and $\mathfrak{a}$ will be 
given by the integral:
\begin{equation}
   w(\mathfrak{a},R_{12})=\int\limits_{\alpha_4}|\Psi(\mathfrak{a},\alpha_4;
   R_{12} )|^2 P(\mathfrak{a},\alpha_4,R_{12})\,d\alpha_4~.
\label{pda}
\end{equation}
The dependence of the fission probability (\ref{pda}) on the mass asymmetry is 
shown in the bottom part of Fig.~\ref{2Dmass} for a few values of $R_{12}$. One 
observes that due to the Fermi function in (\ref{pneck}) the contribution of 
larger $R_{12}$ (smaller necks) is enhanced and contributions from smaller 
$R_{12}$ are suppressed.

From the maps of the potential energy surface shown in the top part of 
Fig.~\ref{evolution} one observes that for $R_{12}\le 2.1$ the neck parameter is 
$\alpha_4=-0.05$ while for  $R_{12}\ge 2.2$ the minimum at the PES is at 
$\alpha_4=-0.15$. This means that a large part of the distribution probability 
$|\Psi(\mathfrak{a},\alpha_4;R_{12}|^2$ will undergo fission also at smaller 
elongations and one has to subtract this part from the phase-space, i.e. to 
diminish the initial distribution by subtracting the events which have already 
fissioned. The final (measured) mass distribution of the fission fragments will 
be the sum of those subtracted events.

\begin{figure}[htb]
\begin{center}
\includegraphics[width=0.35\textwidth]{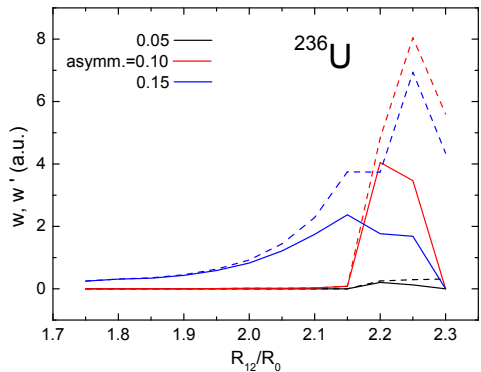}
\end{center}\vspace{-5mm}
\caption{(Color online) Comparison of the fission probabilities
(\protect\ref{waR}) (dashed line) and (\protect\ref{replace}) (solid line) for 
a few values of the mass asymmetry $\mathfrak{a}$=asymm.}
\label{Wi}
\end{figure}

Such an approach means that the fission process should be spread over some 
region of $R_{12}$ and that for given $R_{12}$ at fixed mass asymmetry one has 
to take into account the probability to fission at previous $R_{12}$ points. 
I.e., one has to replace $w(\mathfrak{a},R_{12})$ by
\begin{equation}\label{replace}
w^{\prime}(\mathfrak{a},R_{12})=w(\mathfrak{a},R_{12})
\left(1-\frac{
\int\limits_{R^{\prime}_{12}\le R_{12}}w(\mathfrak{a},R^{\prime}_{12})\,d
R^{\prime}_{12}}{\int\limits^{ } w(\mathfrak{a},R^{\prime}_{12})\,d
R^{\prime}_{12}}\right)\,.
\end{equation}
The effect of the replacement (\ref{replace}) is demonstrated in Fig.~\ref{Wi}.
It is seen there that this replacement substantially reduces the magnitude of 
the fission probability at large $R_{12}$.
%%%%%%%%%%%%%%%%%%%%%%%%%%%%%%%%%%%%%%%%%%%%%%%%%%%%%%%%%%%%%%%%%%%%%%%%%%%%%%%
%%%%%%%%%%%%%%%%%%%%%%%%%%%%%%%%%%%%%%%%%%%%%%%%%%%%%%%%%%%%%%%%%%%%%%%%%%%%%%%

The mass yield will be the sum of all partial yields at different $R_{12}$:
\begin{equation}\label{Ya}
   Y(\mathfrak{a})=\frac{\int\limits_{ } w^{\prime}(\mathfrak{a}, R_{12})\,d
R_{12}} {\int\limits^{ } w^{\prime}(\mathfrak{a}, R_{12})\,d R_{12}\, d
\mathfrak{a}}~.
\end{equation}
As it is seen from (\ref{Ya}) the scaling factor $k_0$ in the expression for 
$P$, Eq.~(\ref{pn}), has vanished and does not appear in the definition of the 
mass yield. Our model will thus only have two adjustable parameters, $\kappa_0$ 
and $d$, that appear in the neck-breaking probability (\ref{pneck}).
%%%%%%%%%%%%%%%%%%%%%%%%%%%%%%%%%%%%%%%%%%%%%%%%%%%%%%%%%%%%%%%%%%%%%%%%%%%%%%%
\begin{figure}[htb]
\begin{center}
\includegraphics[width=0.35\textwidth]{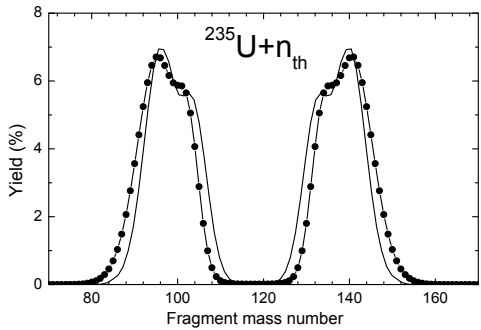}
\end{center}\vspace{-5mm}
\caption{ Comparison of the measured mass distribution of fission fragments 
(points) in the reaction $^{235}{\rm U}+n_{th}$ \protect\cite{exp} with the 
values (\protect\ref{Ya}), calculated with $\kappa_0=0.16, d=0.09$ (solid 
line).}
\label{yields}
\end{figure}
%%%%%%%%%%%%%%%%%%%%%%%%%%%%%%%%%%%%%%%%%%%%%%%%%%%%%%%%%%%%%%%%%%%%%%%%%%%%%%%

A comparison of the measured \protect\cite{exp} and here calculated fission 
fragment mass distributions is shown in Fig.~\ref{yields} for the thermal 
neutron induced fission of $^{235}{\rm U}$.

One can see that the calculated mass distribution is very close to the 
experimental values. The double-peak structure, the position and the relative 
magnitude of the peaks are reproduced rather well.\\

\section{Conclusions}

The extended Cassini ovals deformation parameters and the 
macroscopic-microscopic model ($E_{\rm LD}$ plus the WS single-particle 
potential) yields for $^{236}$U a PES with an asymmetric fission valley 
corresponding to $A_{\rm f}\approx 140$ when the relative distance between the 
fragment mass centers is smaller than $R_{12}=2.3 R_0$. At larger elongations 
one observes a sudden jump of the maximum of the distribution to $A_{\rm 
f}\approx 132$ what causes severe problems in a correct reproduction of the 
data.

We have shown that the three-dimensional quantum mechanical model which couples 
the fission, neck and mass asymmetry modes is able to describe the main features 
of the fragment mass distribution when the neck dependent fission probability is 
taken into account. The obtained mass distribution is slightly shifted, by 
approximately 2 mass units, towards symmetric fission as compared with the 
experimental mass yield, but reproduces nicely the structure of the distribution 
observed in the experiment. This shift could be partly due to a too large 
stiffness of the LD energy in the mass asymmetry degrees of freedom and/or to a 
lack of the nonadiabatic effects (beyond the Born-Oppenheimer app.) \cite{APPBS} 
which makes the distributions slightly wider than the sole adiabatic ones. Also 
the energy dissipation, not taken into account in the present investigation, 
could modify somewhat the theoretical distribution.\\[-5ex]

\begin{acknowledgments}
The authors are very grateful to Drs. Christelle Schmitt and Johann Bartel for 
the careful reading of the manuscript and the valuables comments. 
One of us (F. I.) would like to express his gratitude to the
Theoretical Physics Division of UMCS for the hospitality during his stay at
Lublin. This work has 
been partly supported by the Polish National Science Centre, grant No. 
2013/11/B/ST2/04087.
\end{acknowledgments}

\end{document}